# Diamond optomechanical cavity with a color center for coherent microwave-to-optical quantum interfaces

Byunggi Kim[1,*], Hodaka Kurokawa[2], Katsuta Sakai[3,4], Kazuki Koshino[2,3], Hideo Kosaka[2,5], and Masahiro Nomura[1,*]

[1]*Institute of Industrial Science, The University of Tokyo,*
*4-6-1 Komaba, Meguro, Tokyo 153-8505, Japan*
[2]*Quantum Information Research Center, Institute of Advanced Sciences,*
*Yokohama National University, 79-5 Tokiwadai, Hodogaya, Yokohama 240-8501, Japan*
[3]*College of Liberal Arts and Sciences, Tokyo Medical and Dental University,*
*2-8-30 Konodai, Ichikawa, Chiba 272-0827, Japan*
[4]*KEK Theory Center, Institute of Particle and Nuclear Studies,*
*High Energy Accelerator Research Organization, 1-1 Oho, Tsukuba, Ibaraki 305-0801, Japan*
[5]*Department of Physics, Graduate School of Engineering Science,*
*Yokohama National University, 79-5 Tokiwadai, Hodogaya, Yokohama 240-8501, Japan*

Corresponding author: nomura@iis.u-tokyo.ac.jp, bkim@iis.u-tokyo.ac.jp

**ABSTRACT**

Quantum transduction between microwave and optical photons plays a key role in quantum communications among remote qubits. Although the quantum transduction schemes generating communication photons have been successfully demonstrated by using optomechanical interfaces, the low conversion efficiency remains an obstacle to the implementation of a quantum network consisting of multiple qubits. Here, we present an efficient quantum transduction scheme using a one-dimensional (1D) diamond optomechanical crystal cavity tuned at a color-center emission without optomechanical coupling. The optomechanical crystal cavity incorporates a thin aluminum nitride (AlN) pad piezoelectric coupler near the concentrator cavity region, while retaining ultrasmall mechanical and optical mode-volumes of $\sim 1.5 \times 10^{-4} (\Lambda_p)^3$ and $\sim 0.2 \left(\frac{\lambda}{n}\right)^3$, respectively. The energy level of a coherent color center electron is manipulated by a strong mechanical mode-color-center electron-coupling rate up to 16.4 MHz. In our system, we theoretically predict that the population-conversion efficiency from a single microwave photon into an optical photon can reach 15% combined with current technologies. The coherent conversion efficiency is over 10% with a reasonable pure decay time $T_2^* > 10$ ns. Our results imply that an atomic color center strongly coupled to the optomechanical crystal cavity will offer a highly efficient quantum transduction platform.

**I. Introduction**

Quantum transduction platforms for a single microwave-to-optical conversion are of paramount interest for quantum networks between microwave-controlled qubits. For this purpose, the use of the optomechanical interface [1–6] has emerged as an efficient way of conversion between microwave and optical photons. The scheme using the 1D optomechanical cavity typically integrates a piezoelectric coupler and phononic waveguide [1,2,7]. A microwave photon produces a phonon with the same frequency to couple with the optical photon in optomechanical interfaces via photoelastic and moving-boundary effects [8–10]. The cavity-enhanced interaction between photon and phonon exerts frequency modulation of an optical photon, thereby enabling quantum transduction.

In the last decade, the quantum interface using a diamond color center has attracted enormous attention for the generation of remote entanglement between coherent spin qubits [11–14]. The photonic nanocavity with high cooperativity enables the control of spin and orbital states of the color center electron [15,16], leading to the large-scale integration of multinode quantum processors [17]. In addition, the spin memory-enhanced quantum interfaces have been implemented via the optomechanical system [18]. However, the low microwave-to-optical conversion efficiency remains as a challenge for the implementation of the multinode quantum network. The optomechanical interface requires a significant photonic cooperativity of the optomechanical cavity with a large number of pumping photons [19,20], which generates critical thermal noise in the sub-Kelvin temperatures. To suppress the thermal noise, we have proposed the quantum-interfaces using a diamond spin memory with the photonic cavity-enhanced emission [21]. Our theoretical predictions suggested that the optical pump power decreases by 2-3 orders of magnitude while maintaining the entanglement generation rate of several tens-of-kilohertz.

In this study, we investigated the feasible design of a quantum transduction system enabling microwave-to-optical conversion via a strong mechanical mode-color center electron interaction inside the diamond optomechanical crystal cavity. Specifically, the piezoelectric aluminum nitride (AlN) thin-film pad on the 1D nanobeam optomechanical crystal is used to couple with microwave-emitting qubits, while keeping the ultrasmall mode-volume of the mechanical and optical modes. The mechanical mode-color center interaction enables manipulation of the coherent electron energy level in a charged nitrogen vacancy center (NV⁻) in diamond, followed by emission of an optical photon without the optomechanical coupling. Further, we discuss the performances of the microwave-to-optical conversion by providing the time evolution of the quantum population. The simulation results indicate that a color center electron emits a photon to the optical waveguide with a microwave-to-optical conversion efficiency of 15%. Furthermore, we calculated the microwave-to-optical coherent conversion efficiency as high as 10 % with the pure dephasing time of the NV⁻ $T_2^* = 10$ ns. Hence, we expect that our scheme can be applied to efficient quantum networks with embedded coherent spin-memories.

## II. Results and Discussion

### 1. Quantum interfaces between a superconducting qubit and an optical photon

Figure 1a shows a schematic of the quantum transduction scheme between a superconducting qubit and a cavity photon. Our quantum interfaces include three cavities of microwave photon, phonon, and optical photon. The microwave resonator contains a piezoelectric transducer of a thin-film AlN pad to convert the microwave photon into a phonon, which is excited by the non-contact electrode pair. The generated cavity phonon tunes the energy level of the electron via the interaction between the mechanical mode and the color center electron. In this study, we consider the use of the NV⁻, of which zero-phonon line $\omega_{ZPL}$ is resonant at 470 THz. Use of the NV⁻ center benefits from facile manipulation of the energy level at the excited state [22]. As shown in Fig. 1b, the optical photon excites energy level of the electron at the photonic cavity, with the external optical driving frequency of $\omega_d$. The mechanical mode-color center interaction changes the energy level of the electron by $\hbar\omega_m$, such that $\omega_{ZPL} = \omega_d + \omega_m$. Here, the frequency of the photonic cavity, $\omega_{opt}$, includes both the optical driving, $\omega_d$, and modulated frequencies, $\omega_d + \omega_m$, to satisfy linewidth-limited modulation conditions:

$$\omega_{opt} - \gamma_{opt}/2 < \omega_d < \omega_{opt} + \gamma_{opt}/2 \qquad (1.1)$$

$$\omega_{opt} - \gamma_{opt}/2 < \omega_d + \omega_m < \omega_{opt} + \gamma_{opt}/2 \qquad (1.2)$$

where $\gamma_{opt}$ is the linewidth (decay rate) of the photonic cavity. Thus, the state of the superconducting qubit can be transferred to the cavity photon thorough the optical waveguide.

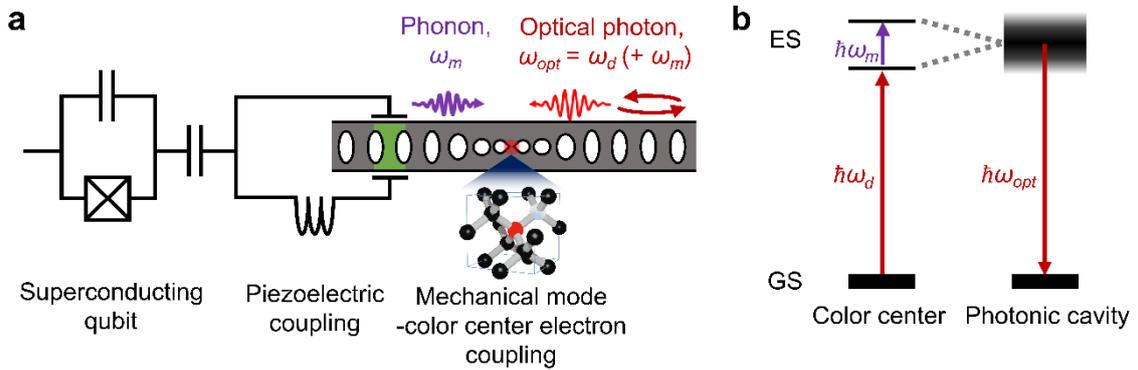

**Figure 1**| Schematic illustration of the microwave-to-optical conversion scheme via a diamond color-center spin memory. **a**, Quantum interfaces to convert a microwave photon generated by a superconducting qubit to the photon tuned at the color-center emission. **b**, The energy level of the color center electron included in that of the photonic cavity.

## 2. Design of the optomechanical cavity

Figure 2a shows the design of the 1D nanobeam system accommodating the microwave resonator and the optomechanical cavity. The optomechanical cavity consists of photonic and phononic mirror cells and 8 cavity cells. The AlN pad is placed on top of the mirror region to convert a microwave photon to a phonon via piezoelectric coupling. The mechanical mode-matching should be realized for efficient coupling between the microwave resonator and optomechanical cavity. In addition, mechanical waveguide loss can be reduced by optimizing the distance between a piezoelectric material and the cavity region of the optomechanical crystal. From these points of view, we considered the use of the small AlN pad near the cavity as a piezoelectric resonator.

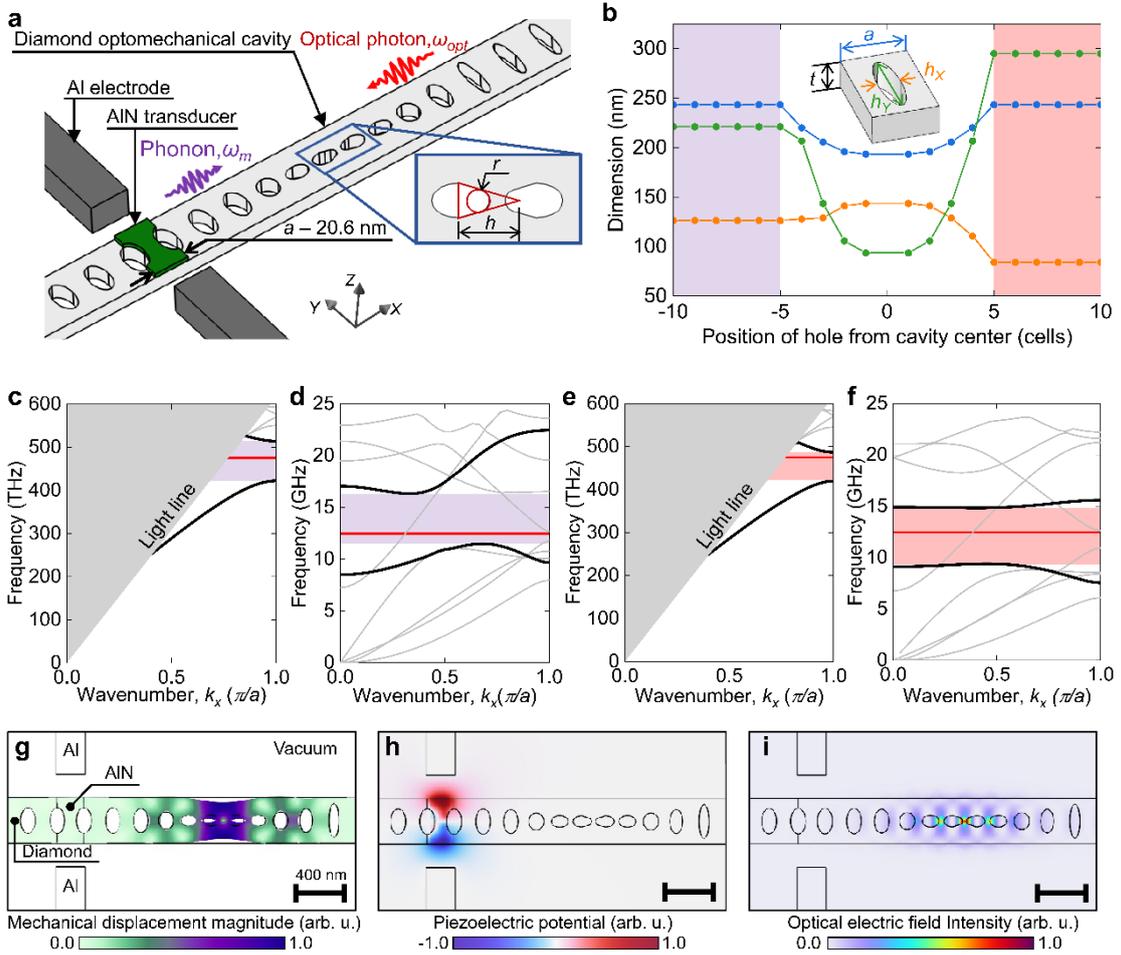

**Figure 2|** Design of the microwave-to-optical quantum transduction system using the 1D optomechanical cavity. **a,** Overview of the diamond optomechanical cavity with the piezoelectric resonator near the cavity. **b,** Dimension of unit cells of the 1D optomechanical cavity. **c, d,** Photonic and phononic band structures of the phononic quasiwaveguide, respectively. **e, f,** Photonic and phononic band structures of the photonic quasiwaveguide, respectively. In **b**, purple (red)-shaded area means phononic (photonic) quasi-waveguide cells, of which band gaps correspond to the same color in **c-f**. **g,** Displacement distribution of the mechanical resonant mode. **h,** Electric potential distribution under the

mechanical resonance. **i**, Electric field distribution of the optical resonant mode.

Cavities with ultrasmall mode-volumes have been extensively investigated for rapid manipulation of electron orbits [23–27]. Specifically, the use of the concentrators at the cavity is simple and robust approach to achieve ultrasmall mode-volumes both for photonic [27–30] and phononic modes [23,26]. The inset of Fig. 2a describes the detailed design of the concentrators. In our cavity design, the radius of curvature at the concentrator tip was set at 30 nm considering a typical fabrication resolution by an electron beam lithography. Figure 2b shows the geometry of the unit cells of the optomechanical cavity. Geometries of the outer half-ellipse are indicated for the central cavity holes. We performed a finite-element method (FEM) simulation to design the optomechanical cavity. The cavity-resonant frequencies are $\omega_{opt}$~470 THz and $\omega_m$~12.5 GHz for photon and phonon, respectively.

The asymmetric design of the mirror allows phononic quasi-waveguiding from the left side, and photonic quasi-waveguiding from the right side, respectively. Figures 2c-f show the photonic and phononic band structures of the left and right sides of the beams. As shaded in Fig. 2b, the band gaps are filled by purple and red for the left and right sides, respectively. For the mechanical vibration, the band gaps of the breathing modes are indicated with the adjacent breathing modes as the black bold lines. The vibrational modes inside the breathing band gaps are orthogonal to the breathing mode due to the differences in symmetry so that no interference occurs. We optimized the unit cell geometries to implement partial mirrors by adjusting the position of the resonant frequencies within the bandgaps. For the left (right) side of the beam, the photonic (phononic) resonant frequency was set in the range of ±15 THz (±1.5 GHz) from the bandgap center to prevent energy leakage. On the other hand, the phononic (photonic) resonant frequency was 1.5 GHz (15 THz) shifted away from the band edge, enabling external input and output. Coupling to the external optical waveguide is realized by changing the number and period of photonic partial mirror cells while maintaining the mechanical quality factor. The cavity and the unit cell period with the AlN pad were optimized to maximize the optical quality factor using the Nelder-Mead method, which has been accepted as one of the efficient ways to design of the 1D optomechanical cavity with several geometrical parameters [7,31,32]. The period of the mirror cell with the AlN transducer was optimized to $a - 20.6$ nm to maximize the optical quality factor.

Figures 2g-i show the FEM simulation results of the mechanical resonant mode, the electric potential field profile under the mechanical resonance, and the optical resonant mode, respectively. The optical mode profile in Fig. 2i was calculated in the presence of the Al to take into account for scattering of the evanescent fields. To maximize the piezoelectric coupling, we considered the deposition of *m*-plane AlN on the diamond slab. Although the growth of the *m*-plane AlN thin film is technologically difficult, remarkable experimental works have been reported by using metalorganic chemical vapor deposition [33] and plasma-nitridation of the *m*-plane sapphire [34]. The mechanical

breathing mode is mainly observed in the cavity region. Accordingly, the electric potential inside the AlN pad increases almost monotonically along the direction perpendicular to the 1D nanobeam. Thus, the microwave excitation can be coupled with the mechanical breathing mode using the electrodes next to the 1D optomechanical crystal.

The mechanical and optical mode-volumes, $V_{mech}$ [8,25] and $V_{opt}$, are given by

$$V_{mech} = \frac{\int_V h(\mathbf{r})\mathrm{d}^3\mathbf{r}}{\max(h(\mathrm{r}))} \tag{2}$$

$$V_{opt} = \frac{\int_V \epsilon(\mathbf{r})|\mathbf{e}(\mathbf{r})|^2\mathrm{d}^3\mathbf{r}}{\epsilon(\mathbf{r}_{max})\max(|\mathbf{e}(\mathbf{r})|^2)} \tag{3}$$

where $\epsilon$, $\mathbf{e}$, $n$, $\lambda$, and $\mathbf{r}$ are permittivity, electric field, refractive index, optical wavelength, and spatial coordinates, respectively. The local energy density $h$ averaged over a period, $2\pi/\omega_m$, is given as a sum of the stored strain and kinetic energy densities:

$$h = \frac{1}{4}[\mathcal{R}e(\sigma:\bar{T}) + \rho\omega_m^2|\mathbf{u}|^2] \tag{4}$$

where $\sigma$, $T$, $\rho$, and $\mathbf{u}$ are stress, strain, density, and mechanical displacement, respectively. The overbar indicates complex conjugate.

Surprisingly, the rounded concentrators mediated ultrasmall mode-volumes of $V_{mech} = 1.5 \times 10^{-4}(\Lambda_p)^3, 5.1 \times 10^{-4}(\Lambda_s)^3$, and $V_{opt} = 0.2(\lambda/n)^3$, where $n$ and $\lambda$ are refractive index and the wavelength, respectively. The longitudinal and shear wavelengths of the mechanical modes are given as $\Lambda_p = 2\pi\sqrt{E(1-\nu)/[\rho(1+\nu)(1-2\nu)]}/\omega_m$ and $\Lambda_s = 2\pi\sqrt{E/[2\rho(1+\nu)]}/\omega_m$, respectively. Here, $E$ and $\nu$ denote Young's modulus and Poisson's ratio, respectively. Our results suggest that a slight asymmetry of the central cavity holes leads to a dramatic reduction of the mode-volume in optomechanical cavities. The mechanical quality factor given by the FEM simulation is in the order of $10^9$, where the loss is solely given by the perfectly matched layers.

Next, we investigated the effects of the positions of the electrodes and the AlN transducer and the distance between the concentrator tips (neck), since the optical mode is sensitive to the scattering losses. Figures 3a and 3b show the optical quality factor $Q_{opt}$ as a function of the geometries of a piezoelectric resonator. The optical quality factor of the optomechanical cavity is 34,000 without the AlN pad. As shown in Fig. 3a, when the AlN pad is positioned more than 9 holes away from the cavity center, the optical quality factor is not affected by the AlN pad. The FEM simulation errors caused the fluctuation of the optical quality factor $Q_{opt}$ around 34,000. Note that optical quality factor of the current diamond 1D diamond nanobeam crystal cavities experimentally reached 42,000 [35] and 1.76×10⁵ [9], which is larger than our design optical quality factor ~12,000. Therefore, the deterioration of the optical quality factor would be insignificant considering use of the state-of-the-art fabrication technologies.

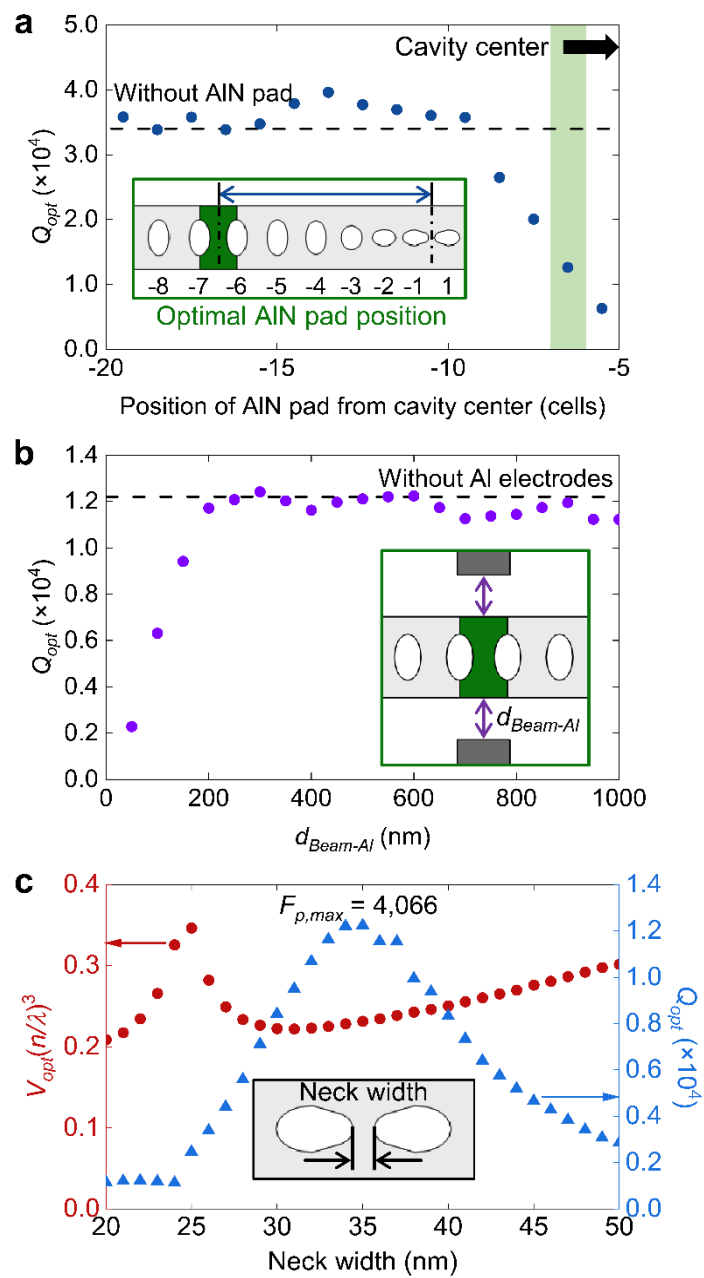

**Figure 3|** Optimization of the photonic cavity. **a**, Optical quality factor vs. position of the AlN pad. The green-shaded region is the optimized AlN pad position. Although the AlN pad should be positioned close to the cavity center for a strong piezoelectric coupling, we set the AlN pad position to keep the optical quality factor over 10,000. **b**, Optical quality factor vs. distance between the 1D optomechanical crystal beam and the electrode, $d_{Beam-Al}$. **c**, Optical mode-volume and quality factor vs. neck width of the concentrators. The insets show the schematic of each geometrical parameter.

For a strong piezoelectric coupling, the AlN pad should be as close as possible to the mechanical cavity to have larger internal electric field. On the other hand, the AlN pad scatters the optical electromagnetic field generated at the photonic cavity. Thus, the position of the AlN pad introduces a trade-off between the piezoelectric coupling and the optical quality factor. In this study, considering the photonic bandwidth to cover the mechanically modulated frequency, we placed the AlN pad between two holes with indices of -6, and -7. For $\omega_{opt}$ ~470 THz, the optical quality factor $Q_{opt}$ of 12,000 has the spectral bandwidth of 39 GHz, which sufficiently covers the microwave modulation of 12.5 GHz. Figure 3b shows that the degradation of the optical quality factor due to scattering of the evanescent field becomes negligible when the distance between the 1D-beam and the electrode, $d_{Beam-Al}$, is larger than 200 nm. In our scheme, since the photonic resonance is at the visible or near-infrared wavelength rather than the communication band (1.5 μm), the width of the 1D nanobeam and $d_{Beam-Al}$ can be reduced to enhance the piezoelectric coupling.

In the concentrator design, most of the electromagnetic field energy is confined within the neck. Therefore, the neck width between the concentrators is an important design parameter for the photonic cavity. Figure 3c shows the optical mode-volume and quality factor as the neck width is varied. Changing the neck width by a few nanometers leads to a dramatic change in the optical quality factor. On the other hand, the optical mode-volume is maintained in the order of $0.2 \sim 0.3 \left(\frac{\lambda}{n}\right)^3$. The ultrasmall mode-volume photonic cavity is able to incorporate an extraordinarily large cooperativity with a deep sub-wavelength structure inside the 1D nanobeam cavity [28–30]. In our design, the maximum Purcell factor was 4,066 at a neck width of 35 nm. In practice, the cooperativity of our system can reach up to 10 to 100, considering the degradation of 1~2 orders of magnitude affected by fabrication imperfection, atomic properties and orientation [15]. The optical resonant frequency is affected by the dimension of the defects, where the energy of the electromagnetic wave is confined. Thus, the neck width changes the optical resonant frequency. As the optical resonant frequency is set at the band edge (Fig. 1e) for implementation of the quasi-waveguide, the neck width has a nontrivial impact on the energy leakage to the photonic waveguide, leading to the variation of the mode-volume appearing at about 25 nm neck width.

## 3. Performances of quantum interfaces

In our previous study [21], we predicted that a significant optical photon generation rate could enable applications such as remote entanglement generation between superconducting qubits. In our scheme, the two quantum interfaces of the piezoelectric coupling and mechanical mode-color-center interaction affect the microwave-to-optical conversion efficiency. The piezoelectric coupling rate can be calculated by the overlap integral between an electric field and an electric displacement field [24,36,37]:

$$g_{MW-m} = \frac{1}{2\hbar} \int_V \left( \mathrm{T}^*(\mathbf{r}) D^T \mathbf{e}(\mathbf{r}) + \mathbf{e}^*(\mathbf{r}) D T(\mathbf{r}) \right) \mathrm{d}^3 \mathbf{r} \tag{5}$$

where $D$ indicates the piezoelectric coupling tensor. Figure 4 shows the microwave photon-phonon coupling in the cavity system. The electric field applied from the side of the 1-D beam (Fig. 4a) leads to the excitation of the mechanical breathing mode (Fig. 4b). Note that the microwave-excited vibrational mode is consistent with the mechanical resonant mode in Fig. 1g. Using Eq. (5), we calculated $g_{MW-m}/(2\pi) = 0.3$ MHz, which is only an order of magnitude smaller than the case of direct electrical excitation near the mechanical cavity reported in [38].

The NV⁻ center has a large mechanical susceptibility $\chi \approx -0.85$ PHz/strain [39] to the strain tensor component of $t_{xx}(r) - t_{yy}(r)$. The coupling rate between the mechanical mode and the color center electron is then given by [23,25]

$$g_{m-e}(\mathbf{r}) = \chi \frac{\left( t_{xx}(\mathbf{r}) - t_{yy}(\mathbf{r}) \right)}{\max(|\mathbf{u}(\mathbf{r})|)} x_{zpf}. \tag{6}$$

The cavity zero-point fluctuation $x_{zpf}$ is given by [23,40,41]

$$x_{zpf} = \sqrt{\frac{\hbar}{2 m_{eff} \omega_m}}, \tag{7}$$

where the effective mass of the resonator is

$$m_{eff} = \frac{\int_V \mathbf{u}^*(\mathbf{r}) \rho(\mathbf{r}) \mathbf{u}(\mathbf{r}) \mathrm{d}^3 \mathbf{r}}{\max(|\mathbf{u}(\mathbf{r})|^2)} \tag{8}$$

Considering use of a diamond (111) slab, we define the high-symmetry axis of the the NV⁻ center (111) is along the $z$-axis of the diamond crystal. Accordingly, the $x$- and $y$-axes are along $(\bar{1}\bar{1}2)$ and $(\bar{1}10)$ directions, respectively.

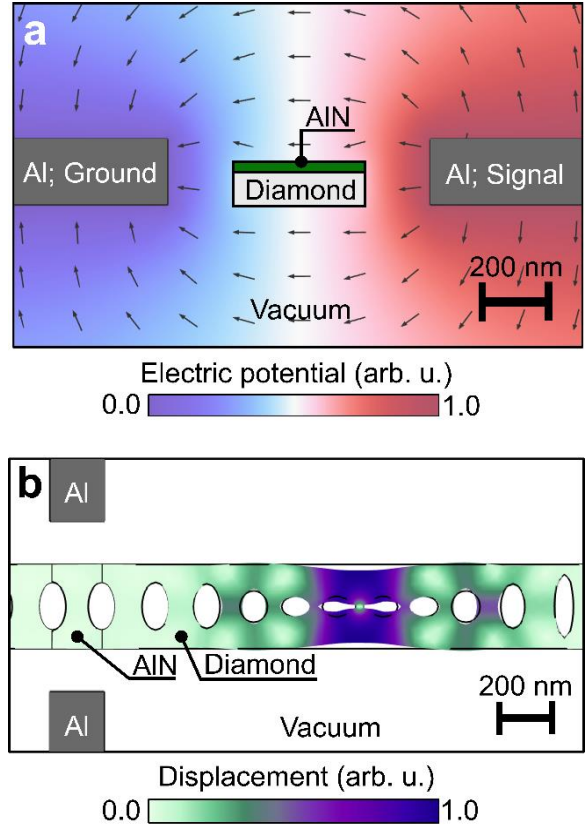

**Figure 4|** Microwave photon-to-phonon conversion via piezoelectric coupling. **a**, Electric potential profile around the 1D optomechanical crystal under application of the microwave electric field. **b**, Resonant mechanical mode of the 1D optomechanical crystal under the microwave electrical field with $\omega_{MW} \sim 12.5$ GHz.

Figure 5 shows the FEM simulation results of the coupling rate between the mechanical mode and the color center $g_{m-e}$ in the mechanical cavity. Raniwala *et al.* [23] showed that the Euler angle between the laboratory coordinate system (*XYZ*) and the crystal coordinate system (*xyz*) has a significant effect on $g_{m-e}$ due to the anisotropy of the 1D-beam. Therefore, we also investigated the effect of the in-plane rotation of the (111) slab as a function of the Euler angle $\phi$ as shown in Fig. 5a. The rotation of the strain tensor is given by

$$T(x, y, z) = RT(X, Y, Z)R^{-1}, \qquad (9)$$

where the rotation matrix $R$ is

$$R = \begin{pmatrix} \cos\phi & -\sin\phi & 0 \\ \sin\phi & \cos\phi & 0 \\ 0 & 0 & 1 \end{pmatrix}. \qquad (10)$$

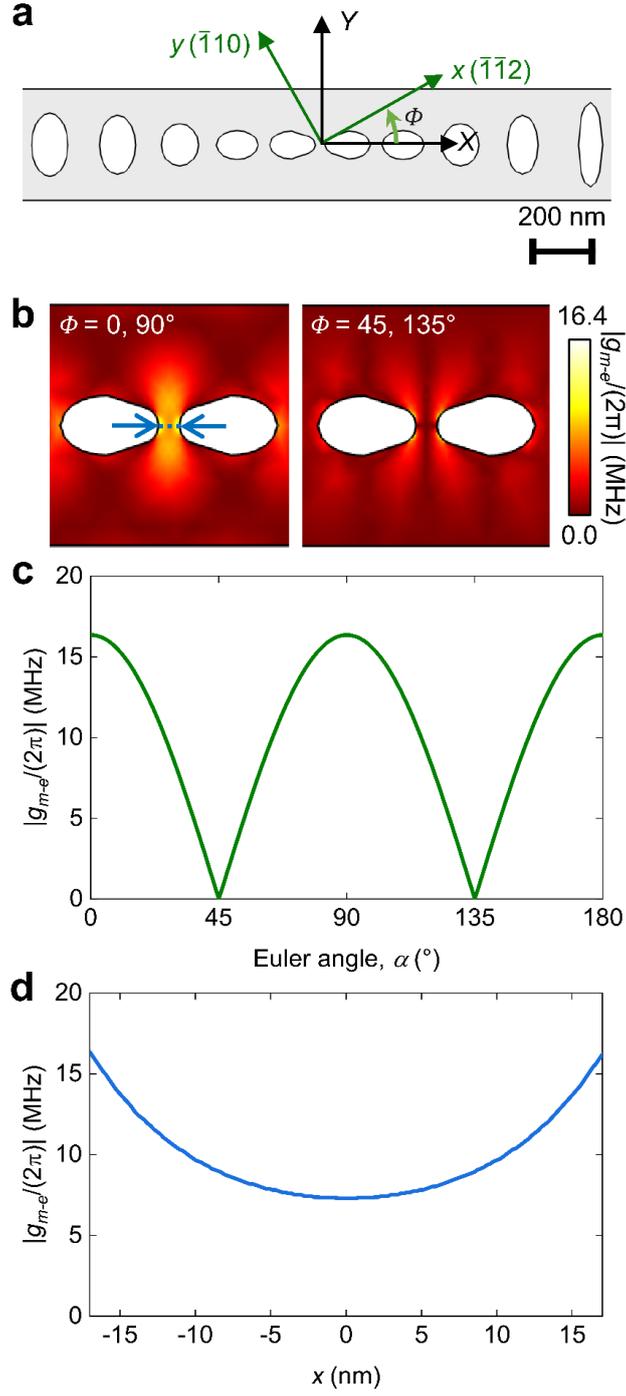

**Figure 5** | Coupling rate between the mechanical mode and the color center electron in the mechanical cavity. **a**, Rotation of the diamond crystal orientation with respect to the 1D optomechanical crystal. **b**, Spatial distribution of $|g_{m-e}/(2\pi)|$ for different Euler angles $\phi$ of $0, 90°$ (left) and $45, 135°$ (right). **c**, $|g_{m-e}/(2\pi)|$ at the concentrator tips for different Euler angles $\phi$ showing the fourfold symmetry. **d**, Spatial profile of $|g_{m-e}/(2\pi)|$ along the neck between the concentrator tips.

In Figs. 5b and 5c, the spatial distribution of $g_{m-e}$ shows a fourfold symmetry, which is originated from the diamond cubic crystal. In particular, the strain tensor component $t_{xx}(r) - t_{yy}(r)$ is maximized when $\phi = 0, 90°$ under the resonant mechanical vibration shown in Fig. 2g. We calculated the maximum $g_{m-e}/(2\pi) = 16.4$ MHz. Figure 5d shows that the minimum $g_{m-e}$ along the concentrator tip-to-tip was 45 % of the maximum value. Therefore, a high mechanical mode-color center coupling rate $g_{m-e}/(2\pi)$ on the order of 10 MHz can be obtained as long as the color center can be placed with the spatial accuracy of 20 nm. Note that the maximum optical mode intensity is homogeneous along the neck (Fig. 2i), thereby mediating the resonant emission of a photon along the neck regardless of the position of the color center.

## 4. Microwave-to-optical conversion efficiency

We estimated the microwave-to-optical conversion efficiency by solving the time evolution of the density matrix, $\rho$, of the quantum interfaces using a quantum toolbox in Python, QuTiP [42,43]. To investigate the coherent microwave-to-optical conversion efficiency, we set the initial state of the microwave as a weak coherent state, which is approximated as $\exp\left(-\frac{|\alpha|^2}{2}\right)(|0\rangle + \alpha|1\rangle)$. The details of the analytical model are described in [21]. Here, we simply present the final forms of the equations. Assuming that the microwave and phonon frequencies are nearly equal to the optical detuning frequency, we can model the Hamiltonian of the quantum interfaces for microwave-to-optical conversion [21,44]:

$$H_{QI,NV^-} = \hbar\omega_{MW}a_{MW}^\dagger a_{MW} + \hbar\omega_m b_m^\dagger b_m + \hbar\Delta_e \sigma_e^+ \sigma_e + \hbar\Delta_{opt} c_{opt}^\dagger c_{opt}$$

$$+ \hbar g_{MW-m}(a_{MW}^\dagger b_m + a_{MW} b_m^\dagger) + \hbar \frac{\Omega_{Rabi} g_{m-e}}{2\omega_m}\left[(b_m^\dagger - b_m)\sigma_e^+ + (b_m - b_m^\dagger)\sigma_e\right]$$

$$+\hbar g_{e-opt}\left\{\left[1 + \frac{g_{m-e}}{\omega_m}(b_m^\dagger - b_m)\right]\sigma_e^+ c_{opt} + \left[1 + \frac{g_{m-e}}{\omega_m}(b_m - b_m^\dagger)\right]\sigma_e c_{opt}^\dagger\right\}, \quad (11)$$

where $\omega_{MW}$, $\Delta_{opt} = \omega_{opt} - \omega_d$ and $\Omega_{Rabi}$ are the frequency of the microwave photon, optical detuning frequency, and optical Rabi frequency, respectively. $a_{MW}^\dagger$ ($a_{MW}$), $b_m^\dagger$ ($b_m$), and $c_{opt}^\dagger$ ($c_{opt}$) are the creation (annihilation) operators of the microwave photon in the piezoelectric resonator, phonon in the optomechanical cavity, and photon in the optomechanical cavity, respectively. $\sigma_e^+$ ($\sigma_e$) is the electron raising (lowering) operator between the ground state and the optically excited state. The master equation in Lindblad form is given by

$$\frac{d\rho}{dt} = \frac{1}{i\hbar}[H_{QI,NV^-}, \rho] + \sum_j \left[\frac{\gamma_j}{2}\left(2c_j\rho c_j^\dagger - c_j^\dagger c_j \rho - \rho c_j^\dagger c_j\right)\right] \quad (12)$$

with $c_j = a_{MW}, b_m, \sigma_e, c_{opt}$ for the energetic decay of each excitation, and $c_j = \sigma_e^\dagger \sigma_e$ for the dephasing of the electron, and $\gamma_j = \gamma_{MW}, \gamma_m, \gamma_e, \gamma_{tot}, \gamma'_e$ being corresponding relaxation rates. In Eq. (12), we have taken into account the coupling to the external photonic waveguide $\gamma_{wg}$ as $\gamma_{tot} = \gamma_{wg} + \gamma_{opt}$, where $\gamma_{opt}$ is the internal loss rate considered as $\omega_{opt}/Q_{opt}$. Here, we set $\gamma_{wg} = \gamma_{opt}$, considering the critical coupling condition between the photonic cavity and the waveguide [45].

We determined the values of the parameters from the FEM simulation results shown in Figs. 2-5 and the literatures: $\omega_{MW}/(2\pi) = \omega_m/(2\pi) = \Delta_{opt}/(2\pi) = 12.5$ GHz, $\Omega_{Rabi}/(2\pi) = 5$ GHz [21], $g_{MW-m}/(2\pi) = 0.3$ MHz (FEM simulation), $g_{m-e}/(2\pi) = 16.4$ MHz, $g_{e-opt}/(2\pi) = 1$ GHz [15], $\gamma_{MW}/(2\pi) = 125$ kHz, $\gamma_m/(2\pi) = 568$ kHz (assuming the microwave photonic quality factor of $10^5$ [46,47], and mechanical quality factor of 22,000 [48]), $\gamma_e/(2\pi) = 10$ MHz. While the FEM simulation gives a significantly large mechanical quality factor of $10^9$, the mechanical quality factor in experiments is limited by internal losses such as phonon-phonon scattering and surface scattering induced by the fabrication imperfection. Recently, the experiment using a silicon nanocavity already achieved a quality factor of ~ $10^{10}$ in the millikelvin [49]. In this study, we adopted the mechanical quality factor of 22,000, which refers to the experiment using a diamond nanocavity under 4 K [48]. We take into account the pure dephasing of NV$^-$ as $\gamma'_e = 1/T_2^*$, where $T_2^*$ is the pure dephasing time. We vary $T_2^*$ in our simulation.

Outputs from the photonic cavity to the waveguide are described by the input-output formalism [50–52] written as

$$d_{out}(t) = d_{in}(t) - i\sqrt{\gamma_{wg}} c_{opt}(t), \qquad (13)$$

where $d_{out}(t)$ and $d_{in}(t)$ are input and output operators of the waveguide, respectively. As our scheme can collect photons emitted from an electron of the NV$^-$, with reflected pumping light discriminated, we can neglect the input operator and treat the output as

$$d_{out}(t) = -i\sqrt{\gamma_{wg}} c_{opt}(t). \qquad (14)$$

The population of photons in the waveguide is given by the time integral of the expectation value of the number operator, $\langle d_{out}^\dagger d_{out} \rangle$. Hence, we define the microwave-to-optical population conversion efficiency $\eta_{pop}$ as

$$\eta_{pop} = \frac{\int_0^{t_f} \langle d_{out}^\dagger d_{out} \rangle dt}{\langle a_{MW}^\dagger a_{MW} \rangle_{t=0}} = \frac{\gamma_{wg} \int_0^{t_f} \langle c_{opt}^\dagger c_{opt} \rangle dt}{\langle a_{MW}^\dagger a_{MW} \rangle_{t=0}},$$

(15)

where the measuring time $t_f$ is large enough. Similarly, by considering off-diagonal elements of the density matrix as the coherence of the system, the microwave-to-optical coherent conversion efficiency $\eta_{coh}$ is given by

$$\eta_{coh} = \frac{\int_0^{t_f}|\langle d_{out}\rangle|^2 dt}{|\langle a_{MW}\rangle|^2_{t=0}} = \frac{\gamma_{wg}\int_0^{t_f}|\langle c_{opt}\rangle|^2 dt}{|\langle a_{MW}\rangle|^2_{t=0}}.$$

(16)

See **Appendix** for the details of their definitions.

First, we simulate an ideal conversion process, *i.e.*, we evaluate the population conversion efficiency with $T_2^* = \infty$. Figure 6 shows the populations of the quantum interfaces and $\eta_{pop}$ as a function of quality parameters of quantum interfaces. In Fig.6a, the population of the optical waveguide photon corresponds to the microwave-to-optical conversion efficiency ~15%. A strong electro-optical coupling $g_{e-opt}/2\pi$ expedites generation of an optical photon, leading to a low population of the orbital excited state of the NV⁻. Furthermore, since the NV⁻ has a large strain susceptibility, the population of the waveguide photon is significantly increased by a large mechanical mode-color center electron coupling rate $g_{m-e}$ at the mechanical cavity with the ultrasmall mode-volume. However, improving $g_{m-e}$ is limited by the fabrication resolution and the trade-off between the piezoelectric coupling rate $g_{MW-m}$ and the optical quality factor $Q_{opt}$ as shown in Fig. 3. Accordingly, the enhancement of the microwave-to-optical conversion efficiency relies on the mechanical quality factor $Q_m$ and the microwave quality factor $Q_{MW}$. Figure 6b shows the effect of the mechanical quality factor and the microwave quality factor on $\eta_{pop}$. High conversion efficiency in the range of 15~35% can be obtained with a moderate improvement in the $Q_m$ and $Q_{MW}$ from the condition used in Fig. 6a, which is indicated as a star.

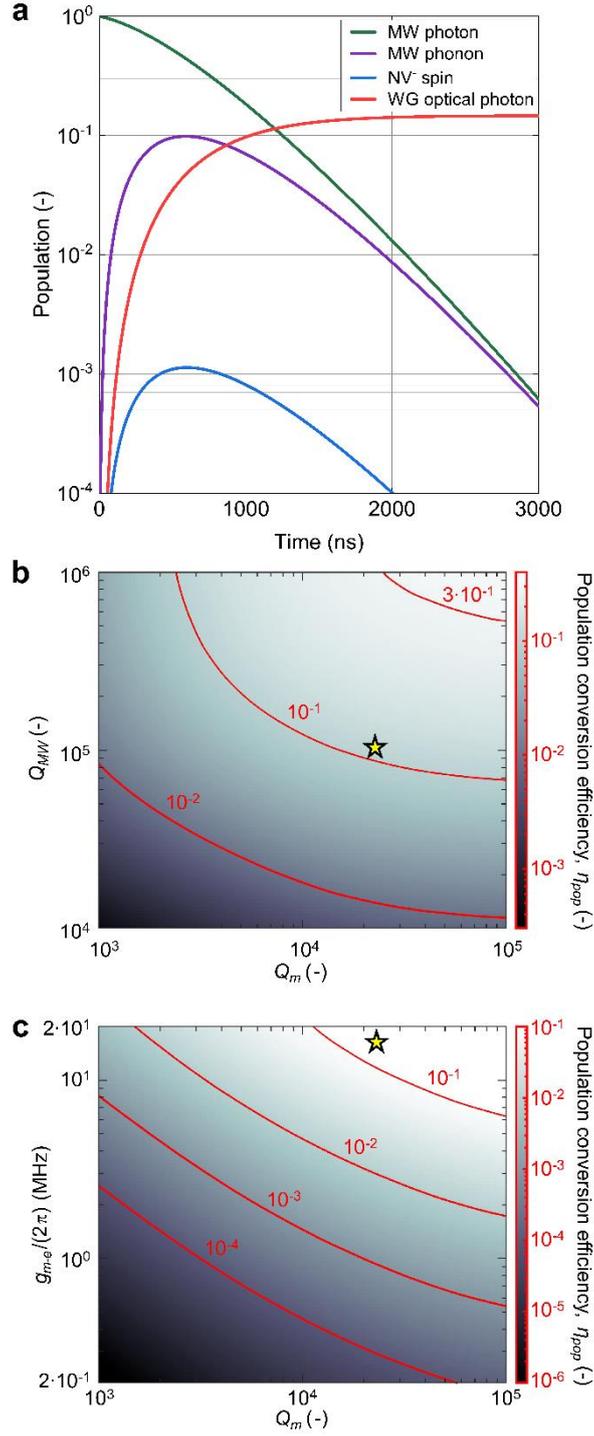

**Figure 6**| Population of quanta. **a**, Time-evolution of the population inside the quantum interface considering parameters obtained from the FEM simulations and the state-of-the-art technologies. **b**, Population conversion efficiency $\eta_{pop}$ as a function of the microwave cavity quality factor $Q_{MW}$ and mechanical cavity quality factor $Q_m$. **c**, $\eta_{pop}$ as a function of the mechanical mode-color center electron coupling rate $g_{m-e}$ and $Q_m$. A star denotes the condition used in **a**: $(Q_{MW}, g_{m-e}/(2\pi), Q_m) = (10^5, 16.4\text{ MHz}, 2.2 \times 10^4)$.

Nanofabrication of diamond inevitably introduces spectral diffusion of color centers due to defect charges and static strains [53]. Meesala *et al.* [54] showed that the strain susceptibility of a silicon vacancy center decreases with the presence of the static strain, thereby decreasing $g_{m-e}$ given by Eq. (6). Figure 6c indicates $\eta_{pop}$ as a function of $g_{m-e}$ and $Q_m$ to investigate effect of rough side walls introduced by nanofabrication processes. As $Q_m$ over $10^4$ has already been demonstrated in the literature [48], achieving $g_{m-e}/(2\pi)$ larger than 10 MHz is an important challenge for the high population conversion efficiency $\eta_{pop} > 10\%$. Meanwhile, a recent investigation [53] successfully reduced surface damage on diamond nanopillars, showing the long-term linewidth stability of the excited state of NV center as low as 150 MHz. Therefore, we expect that advances of fabrication technologies lead to the efficient realization of our scheme in the near future.

Figure 7 shows $\eta_{pop}$ and $\eta_{coh}$ as a function of $T_2^*$, with an amplitude of the weak coherent initial state $\alpha = 0.1$. With increasing values of $T_2^*$, $\eta_{coh}$ asymptotically increased to $\eta_{pop}$, and exceeded 10% for $T_2^*$ larger than 10 ns. Experimentally reported values of $T_2^*$ for NV$^-$ is in the range of 10~80 ns [55]. Thus, the coupling between the NV$^-$ center and the photonic cavity $g_{e-opt}/(2\pi)$ is one-to-two orders of magnitude larger than the reported pure dephasing rate $\gamma'_e$, implying that our scheme facilitates the coherent microwave-to-optical conversion. We have also checked that these results are independent of $\alpha$ as long as $|\alpha|^2 \ll 1$, which is a natural consequence with respect to the linear response.

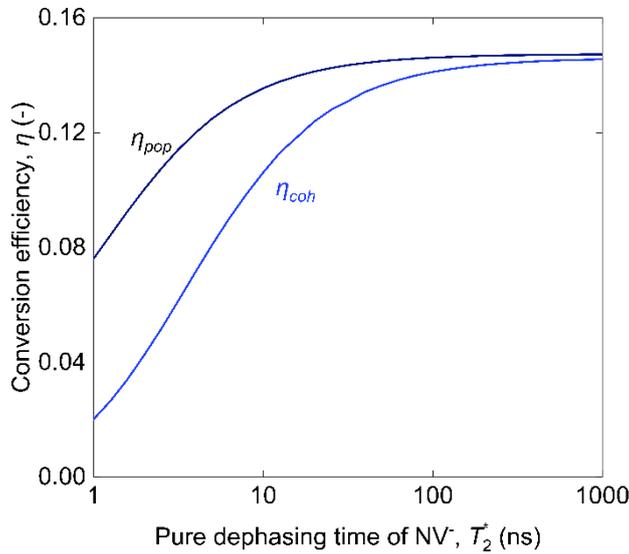

**Figure 7**| Microwave-to-optical population and coherent conversion efficiencies as a function of pure dephasing time of NV$^-$. The microwave initial state was set as $|0\rangle + \alpha|1\rangle$ with $\alpha = 0.1$.

Table 1 summarizes current demonstrations of the microwave-to-optical transduction using a piezo-optomechanical transducer along with design values of this study. It is worth noting that direct comparison is difficult as a low microwave-to-mechanical efficiency was limiting the total conversion efficiency for several literatures [2,6,56]. Overall performances of the microwave-to-optical quantum interfaces have advanced for various platforms incorporating strong microwave-to-mechanical and optomechanical couplings. Our study intimates that a diamond can be a good candidate platform with atomic defect strongly bridging mechanical and optical frequencies. In addition, since our spin memory-based scheme reduces the optical pump power [21], the mechanical quality factor may be further increased from the result shown in ref. [48], by decreasing localized phonon-phonon scattering events in nanostructures. Also, a microwave resonator with an internal impedance converter realized a high quality factor over $10^5$, paving a way for future progress of fast and active control of superconducting qubits [46]. Therefore, implementation of our design will lead to realization of an unprecedentedly high microwave-to-optical quantum converter, combined with the state-of-the-art technologies of nanofabrication and microwave resonator design.

Table 1| Comparison of piezo-optomechanical transducers for MW-to-optical conversion

| References | Platform | $\omega_{MW}/(2\pi)$ (GHz) | $\omega_{opt}/(2\pi)$ (THz) | $g_{om}/(2\pi)$ (Hz)* | $g_{m-e}/(2\pi)$ (Hz) | $\eta_{pop}$ |
|---|---|---|---|---|---|---|
| [57] | AlN | 3.8 | 197 | $1.1 \times 10^5$ | - | $9 \times 10^{-8}$ |
| [6] | GaAs | 2.7 | 194 | $1.3 \times 10^6$ | - | $5.5 \times 10^{-12}$ |
| [58] | LN** | 1.85 | 195 | $8 \times 10^4$ | - | $1.1 \times 10^{-5}$ |
| [59] | AlN | 10 | 200 | $1.9 \times 10^4$ | - | $7.3 \times 10^{-4}$ |
| [1] | AlN on Si | 5.2 | 194 | $7 \times 10^5$ | - | $8.8 \times 10^{-6}$ |
| [2] | GaP*** | 3.2 | 193 | $2.9 \times 10^5$ | - | $1.4 \times 10^{-11}$ |
| [56] | GaP | 2.8 | 193 | $7 \times 10^5$ | - | $6.8 \times 10^{-8}$ |
| [60] | LN on Si | 3.6 | 194 | $4.1 \times 10^5$ | - | $2.5 \times 10^{-2}$ |
| This work | Diamond | 12.5 | 470 | - | $1.6 \times 10^7$ | $1.5 \times 10^{-1}$ |

*Optomechanical coupling rate

**Lithium niobate

***Gallium phosphide

III. Conclusion

We proposed the practical design of the quantum interfaces using the photonic cavity at the color center emission for the quantum transduction between microwave and optical photons via diamond spin-memories. The pair of non-contact electrodes with the 1D optomechanical crystal cavity could generate a phonon by the piezoelectric coupling in the AlN thin film pad to achieve a reasonable piezoelectric coupling rate of 0.3 MHz. By adopting the ultrasmall mode-volume cavity with the rounded concentrators, we calculated the coupling rate $g_{m-e}/(2\pi) = 16.4$ MHz which is one-to-two orders of magnitude larger than typical values, subsequently accelerating emission to the photonic waveguide with the system population conversion efficiency ~15%. Our results imply that an atomic defect coupled to the photonic cavity serves as a coherent quantum transducer, which we predict a coherent conversion efficiency of over 10%. We can also consider an alternative color center with a large strain susceptivity, such as the ground state of $SiV^-$ [61,62]. While our scheme generates non-communication band photons, the on-chip nonlinear photonic platform using silicon carbide [63,64] can be effectively used to convert optical frequencies to extend the distance range of the quantum network. Since our system can provide solutions to practical problems, such as conversion efficiency and thermal noise suppression, we expect that the experimental demonstration will open an alternative pathway for the realization of the millions-node quantum repeaters.

**Data availability**

All data are available from the corresponding authors upon reasonable request. The COMSOL simulation file (Figs.2-5) and Python script (Figs.6-7) are available from the public repository [65].

**Acknowledgements**

Authors appreciate M. Yamamoto and Y. Sekiguchi for insightful discussions. This work was supported by Japan Science and Technology Agency Moonshot R&D grant (JPMJMS2062) and by the Japan Society for the Promotion of Science Grants-in-Aid for Scientific Research (21H04635).

**Appendix: Definition of conversion efficiencies**

Here we address the definition of $\eta_{pop}$ and $\eta_{coh}$ in detail. In a general system, we consider a process where an excitation of one degree of freedom (DoF), denoted by $A$, is transmitted to other DoFs, $B_k$ in time $t_f$. Our aim is to obtain the time evolution of a single photon input $|1\rangle_A |E\rangle$, where $|E\rangle$ is the initial environment state. The process is represented as

$$
\begin{aligned}
|1\rangle_A|E\rangle \rightarrow &\sum_k C_k(t_f)|1\rangle_{B_k}|E\rangle \\
&+\sum_k D_k(t_f)|1\rangle_{B_k}|E'_k\rangle \\
&+\sum_F D'_F(t_f)|0\rangle|F\rangle,
\end{aligned} \quad (17)
$$

where $|E'_k\rangle$ is another state which has the same energy as $|E\rangle$ has, and $|F\rangle$ is a state with a single excitation added. The second term in the RHS of Eq. (17) represents the coherent transmission of the excitation, while the third term expresses the incoherent transmission. The last term denotes losses of the excitation to the environment.

Assuming that we can detect all the excitations of $B_k$, we can define the efficiency of the total transmission of the population as

$$\eta_{pop} = \sum_k |C_k(t_f)|^2 + \sum_k |D_k(t_f)|^2, \quad (18)$$

and the efficiency of the coherent emission as

$$\eta_{coh} = \sum_k |C_k(t_f)|^2. \quad (19)$$

For numerically calculating the conversion efficiencies, we can set the initial state as $\alpha|0\rangle + \beta|1\rangle_A$ with $\alpha^2 + \beta^2 = 1$. In terms of correlation functions, we can rewrite Eqs. (18) and (19) in the following form:

$$\eta_{pop} = \frac{\sum_k \langle b_k^\dagger b_k \rangle_{t=t_f}}{\langle a^\dagger a \rangle_{t=0}}, \quad \eta_{coh} = \frac{\sum_k |\langle b_k \rangle|^2_{t=t_f}}{|\langle a \rangle|^2_{t=0}}, \tag{20}$$

where $a$ and $b_k$ is the annihilation operators for $A$ and $B_k$.

For the model we analyze in the present paper, the output DoFs correspond to the electric field excited on each spatial point of the waveguide. That is, $b_k \to d_r$ with $r$ denoting the position on the waveguide. Our device is connected to the waveguide at $r = 0$. Experimentally, we can collect all the excitations since they propagate to the detector in order. Note that the coefficients $C_r(t_f)$ and $D_r(t_f)$ is nonzero only when $r < t_f$ holds, with the phase velocity is normalized to unity. It is due to the linear propagation of the electric field

$$d_r(t) = d_{out}(t - r) \tag{21}$$

and an assumption that the waveguide initial state is the ground state. By using the above replacements on Eq. (20), and noting that $\sum_k \ldots$ corresponds to $\int_0^{t_f} dr \ldots$, which is equivalent to $\int_0^{t_f} dt \ldots$, we obtain the expression we present in Eqs. (15) and (16).